\documentclass[conference]{IEEEtran}

\usepackage[utf8]{inputenc}
\usepackage[T1]{fontenc}
\usepackage{cite}
\usepackage{amsmath,amssymb,amsfonts}
\usepackage{algorithmic}
\usepackage{graphicx}
\usepackage{textcomp}
\usepackage{xcolor}
\usepackage{balance}

\usepackage{lipsum}
\usepackage{xspace}
\usepackage{hyperref}

\usepackage{color}
\usepackage{lipsum}             
\usepackage{xargs}   
\usepackage{subfigure}
\usepackage{booktabs}
\usepackage{colortbl}

\definecolor{dkgreen}{rgb}{0,0.6,0}
\definecolor{gray}{rgb}{0.5,0.5,0.5}
\definecolor{mauve}{rgb}{0.58,0,0.82}

\newcommand{\eg}{{\textit{e.g.,}}}

\newcommand{\al}{\textit{et al.}\xspace}

\newcommandx{\textblue}[1]{\textcolor{blue}{#1}}

\AtBeginDocument{%
  \providecommand\BibTeX{{%
    \normalfont B\kern-0.5em{\scshape i\kern-0.25em b}\kern-0.8em\TeX}}}

\begin{document}

\title{DR-Tools: a suite of lightweight open-source tools to measure and visualize Java source code}
%% Rever titulo, mas ultima coisa
 
\author{\IEEEauthorblockN{Guilherme Lacerda$^{1,3}$}
\IEEEauthorblockA{$^1$\textit{Unisinos}\\
São Leopoldo, Brazil\\
guilhermeslacerda@gmail.com}
\and
\IEEEauthorblockN{Fabio Petrillo$^{2}$}
\IEEEauthorblockA{$^2$\textit{Universite du Quebec a Chicoutimi}\\
Chicoutimi, Canada\\
fabio@petrillo.com}
\and
\IEEEauthorblockN{Marcelo S. Pimenta$^{3}$}
\IEEEauthorblockA{$^3$\textit{Universidade Federal do Rio Grande do Sul}\\
Porto Alegre, Brazil\\
mpimenta@inf.ufrgs.br}
}

\maketitle

\begin{abstract}
In Software Engineering, some of the most critical activities are maintenance and evolution. However, to perform both with quality, minimizing impacts and risks, developers need to analyze and identify where the main problems come from previously. In this paper, we introduce \textit{DR-Tools Suite}, a set of lightweight open-source tools that analyze and calculate source code metrics, allowing developers to visualize the results in different formats and graphs. Also, we define a set of heuristics to help the code analysis. We conducted two case studies (one academic and one industrial) to collect feedback on the tools suite, on how we will evolve the tools, as well as insights to develop new tools that support developers in their daily work.

Videos: \url{https://bit.ly/30weexX}


\end{abstract}

\section{Introduction}
\label{sec:introduction}

%% Introdução do tema
Metrics have been studied and explored in several studies along many years \cite{Nunez:2017}.
Several studies explore metrics in different contexts like complexity \cite{McCabe:1976}, object-oriented programming \cite{Lanza:2010}, cohesion/coupling \cite{Bob:2007}, architecture \cite{Koziolek:2011}, among others. 
The metrics are also used as an approach to smell detection \cite{Lacerda:2020}, fault prediction \cite{Radjenovic:2013}, and managing the technical debt \cite{Charalampidou:2019}. 
There are still problems with understanding and interpreting the metrics, and which metrics are used in actual situations \cite{Bouwers:2013}.
Some problems are related to naming inconsistencies (same names and different meanings for the same metric) and similar meanings (different names for the same metric), discussed in more detail in \cite{Saraiva:2015}.
Likewise, there are numerous metrics tools, with different features and resources (to find a list of tools, see \cite{Nunez:2017, Fregnan:2019}). Some studies have conducted comparative assessments of several metrics tools \cite{Kayarvizhy:2016, Mshelia:2017}.
Metrics tools with visualization features have also been the subject of previous studies \cite{Systa:2000, Winter:2013}. According to Fregnan \al \cite{Fregnan:2019}, the visualization feature is still a trend, along with extensibility and scalability.
However, just presenting the metrics or even providing visualization mechanisms without having a way to facilitate the developer analysis process or relate to the workflow makes it difficult to use. A study on adopting static analysis tools \cite{Sadowski:2018} showed that simple tools, connected to the workflow, with filtering features \cite{Sadowski:2015} are the most interesting for the developer.

In this paper, we introduce the \textit{DR-Tools Suite}, a lightweight open-source tools that help the developer understand the software.
\textit{DR-Tools Suite} is organized as two tools: (1) \textit{DR-Tools Metric}, a Command-Line Interface (CLI) tool that collects and shows different source code metrics (project summary, namespaces, types, methods, dependencies, and coupling), combined a contextually sorted, and (2) \textit{DR-Tools Metric Visualization}, a tool that generates data visualization in different graphical formats from data generated by \textit{DR-Tools Metric}.
\textit{DR-Tools Suite} is lightweight, designed to be independent of environments and platforms, facilitating interoperability. Its open architecture allows both the functionalities and the resulting data in known standardized formats can be integrated with other tools already adopted by the team, without additional installation or configuration.
We use a well-known set of combined metrics from software metrics research. Also, we define a set of heuristics for the combination of metric based on relationships and thresholds, making use of inferences on the source code. The tools aim to retrieve source code information, providing insights to help developers learn about software complexity and how to improve software maintenance and evolution. Our goal is to help to reduce the cognitive overload of developers when analyzing the results provided by \textit{DR-Tools Suite}. This paper is outlined as follows: Section \ref{sec:drtools-suite} shows the \textit{DR-Tools Suite} concepts, as well as the tools of metrics and visualization. Then, Section \ref{sec:usageexamples} demonstrates two real cases where the tools were used. Section \ref{sec:discussion} shows the user's opinion and a discussion about the issues that were raised. Section \ref{sec:conclusion} contains the conclusion and future works.

\section{DR-Tools Suite}
\label{sec:drtools-suite}

\textit{DR-Tools Suite}\footnote{Information about tools, downloads, complementary materials, metric thresholds used, GitHub link, and others in \cite{drtools:2020}} is a set of lightweight open-source tools\footnote{Tools are under MIT License} that provide resources and information to improve source code quality, supporting the developer in his daily work. 
\textit{DR-Tools Suite} was inspired by the \textit{medicine metaphor} \cite{Feathers:2004}.

\subsection{Selecting a metric set}

Some studies, like Radjenovic et al. \cite{Radjenovic:2013}, show that the relationship between metrics is fundamental for understanding the source code. For example, the relation between size metrics and OO metrics help analyze aspects of code maintainability.
According to Bigonha \al \cite{Bigonha:2019}, when a metric is associated with some threshold, it facilitates its use and understanding. Besides, an isolated metric does not provide much information \cite{Pantiuchina:2018}. In this sense, we create a mechanism to observe metrics used on real situations that can work together and that, ordered with based on a criterion, would allow the developer to infer analysis of the code. \textit{DR-Tools Suite} is the practical implementation of this approach. Based on the contextual selected metrics set, we define some heuristics (Table \ref{tab:heuristics}), which combine metrics and thresholds. 
Through these metrics and heuristics, the developer can identify and understand parts of the code that require more attention according to the context and prioritize actions with the team. Next, we present the tools in detail.

\renewcommand{\arraystretch}{2}
\newcommand{\myrowcolour}{\rowcolor[gray]{0.925}}

\begin{table}[ht]
	\caption{Examples of Analysis Heuristics - The complete list can be found in \cite{drtools:2020}}
	\label{tab:heuristics}
	\scriptsize
	\begin{center}
	   \tiny{
		\begin{tabular}{p{0.2cm} p{0.5cm} p{2.8cm} p{3.5cm} }
			\toprule
			\textbf{\#} & \textbf{Context} & \textbf{Heuristic} & \textbf{Observation} \\
			\hline
		    1 & Namespace & Observe the distribution of classes by namespace & If a namespace has many classes (high NOC), it can be indicative of \textit{promiscuous package}  \\

		    \myrowcolour
            2 & Type & Evaluate metrics beyond the SLOC & WMC, DEPS (DEP and I-DEP) and NOM/NPM are good indications of how the class is doing  \\
		    
		    3 & Type & High SLOC, but without many methods (low NOM/NPM)  & It may be indicative of \textit{long methods}  \\ 
		    
		    \myrowcolour
            4 & Type & High SLOC and WMC, but without many methods (low NOM/NPM) & It can be indicative of \textit{complex class}  \\
		    
		    5 & Method & High NBD can be a complex/long method & Indicative of complexity, legibility and understanding problem \\

		    \myrowcolour
            6 & Coupling & High CE may indicate that the namespace is unstable & The incidence of change in other namespaces that this namespace depends on will cause it to change \\
            
            \bottomrule
		\end{tabular}
		}
	\end{center}
\end{table}

\subsection{DR-Tools Metric}
\label{sec:drtools-metric}

\textit{DR-Tools Metric} is a CLI tool based on a metric set that analyzes the source code, generating results in different formats (pretty format, CSV, and JSON) for each project context.
To use \textit{DR-Tools Metric}, it is not necessary any configuration or installation of any complementary software or plug-in\footnote{\textit{Java Runtime Environment} 8 or higher is required}. \textit{DR-Tools Metric} presents the 33 metrics contextualized by project summary, namespaces (packages), types (classes), methods, dependencies, and coupling (namespace and type).
The following is the list of metrics by context:\\
\textbf{Summary (9):} Total of namespaces, total of types, mean number of types/namespaces, total of lines of code\footnote{Blanks and comment lines are dismissed in all LOC computations} (SLOC), mean number of SLOC/types, total of methods, mean number of methods/types, total of complexity (CYCLO), and mean number of complexity/types;\\
\textbf{Namespaces (2):} Number of classes/types (NOC) and number of abstract classes (NAC);\\
\textbf{Types (9):} Lines of code (SLOC), number of methods (NOM), number of public methods (NPM), class complexity (WMC), number of dependencies\footnote{frameworks and external libraries} (DEP), number of internal dependencies\footnote{Only project classes} (I-DEP), number of other types that depend on a given type (FAN-IN), number of other types referenced by a type (FAN-OUT), and number of fields/attributes (NOA);\\
\textbf{Methods (5):} Lines of code (MLOC), cyclomatic complexity (CYCLO), number of invocations (CALLS), nested block depth (NBD), and number of parameters (PARAM); \\
\textbf{Namespace Coupling (5):} Afferent coupling (CA), efferent coupling (CE), instability (I), abstractness degree (A), and normalized distance (D); \\
\textbf{Type Coupling (4):} number of dependencies (DEP), number of internal dependencies (I-DEP), number of other types that depend on a given type (FAN-IN), and number of other types referenced by a type (FAN-OUT); \\
\textbf{Dependencies (3):} General dependencies (DEP), internal dependencies (I-DEP), and cyclic dependencies. 

\begin{figure}[ht]
\centering
\includegraphics[width=1\linewidth]{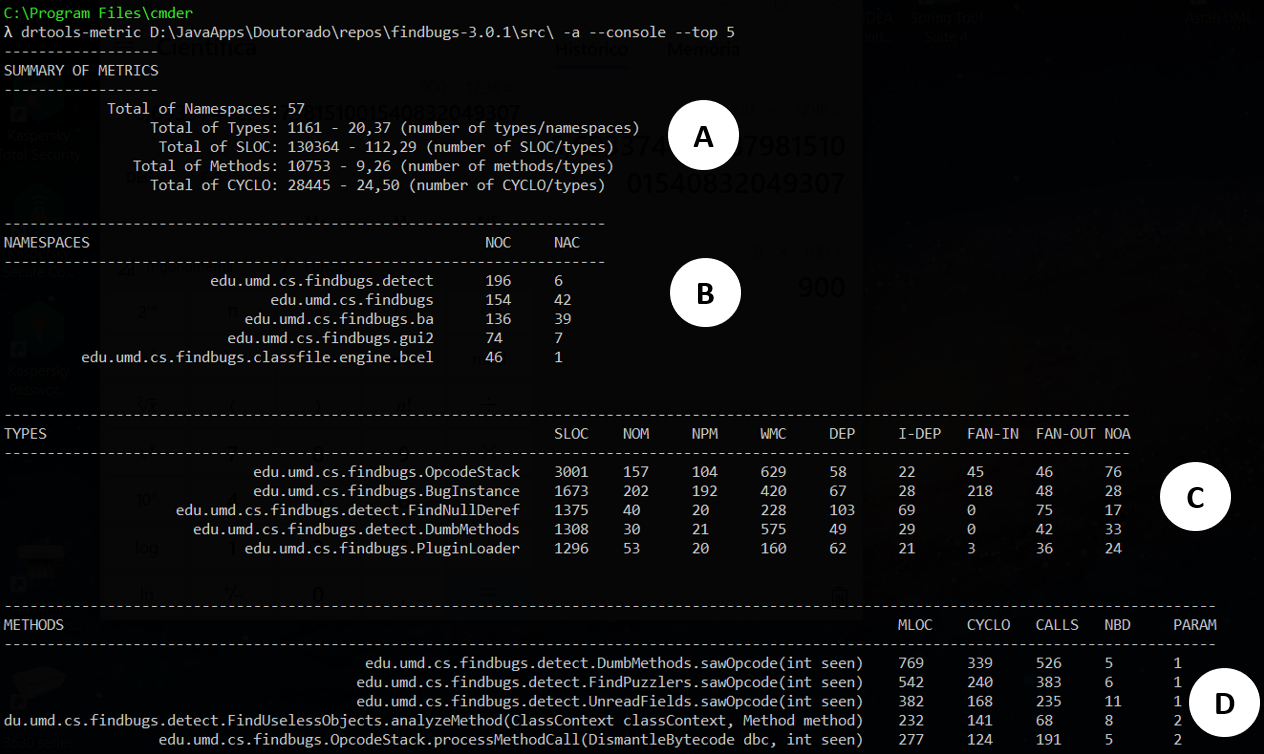}
\caption{\textit{DR-Tools Metric} with the \textit{-a} option, listing top 5 information about summary (A), namespaces (B), types (C), and methods (D) from \textit{FindBugs}. The analysis heuristics (Table I) help in the investigation process. For example, when analyzing packages (B), we notice that the \textit{edu.umd.cs.fingbugs.detect} has many classes and that it can be indicative and that it needs to be better distributed (heuristic 1). In the context of methods, \textit{sawOpCode(int seen)} is the most complex and possibly an inherited method. Choosing the method with less MLOC (\textit{UnreadFields.sawOpCode(int seen)}), but with a high NBD, it is indicative of a complex method (heuristic 5). This information can be confirmed by analyzing the source code.}
\label{fig:drtools-metric-all}
\end{figure}

\textit{DR-Tools Metric} allows developers to combine and query contextual information, from general information (summary), information about packages, classes, methods, dependency types, couplings, and reference thresholds of metrics.
When presenting the results, the data are sorted according to the context. For example, when information about classes, data is sorted by lines of code, complexity, and number of methods. In the context of the method, on the other hand, the combination is cyclomatic complexity, nested blocks, lines code, and invocations. It is also possible to filter contextualized results using the \textit{--top} option. Like this, it's easier for developers to analyze the source code and filter out the most problematic elements. As presented in Figure \ref{fig:drtools-metric-all}, it is possible to have a view on summary and packages, more complex classes, and methods (showing the first 5), in a single option.

\subsection{DR-Tools Metric Visualization}
\label{sec:drtools-metric-visualization}

We developed a tool to view the data generated by \textit{DR-Tools Metric}. 
With a simple \textit{web server}, the developer can run \textit{DR-Tools Metric Visualization}\footnote{Online demo is available in \url{https://metric-visualization-demo.herokuapp.com/}}.
To use the generated data, the developer creates a folder of your project within the \textit{datasets} folder. Then, the developer must copy the generated files (CSV and JSON) to the created folder. Next, the developer needs to edit the \textit{drtools-properties.js} file, redirecting to the project he wants to view the information.

\begin{figure}[ht]
\centering
\includegraphics[width=1\linewidth]{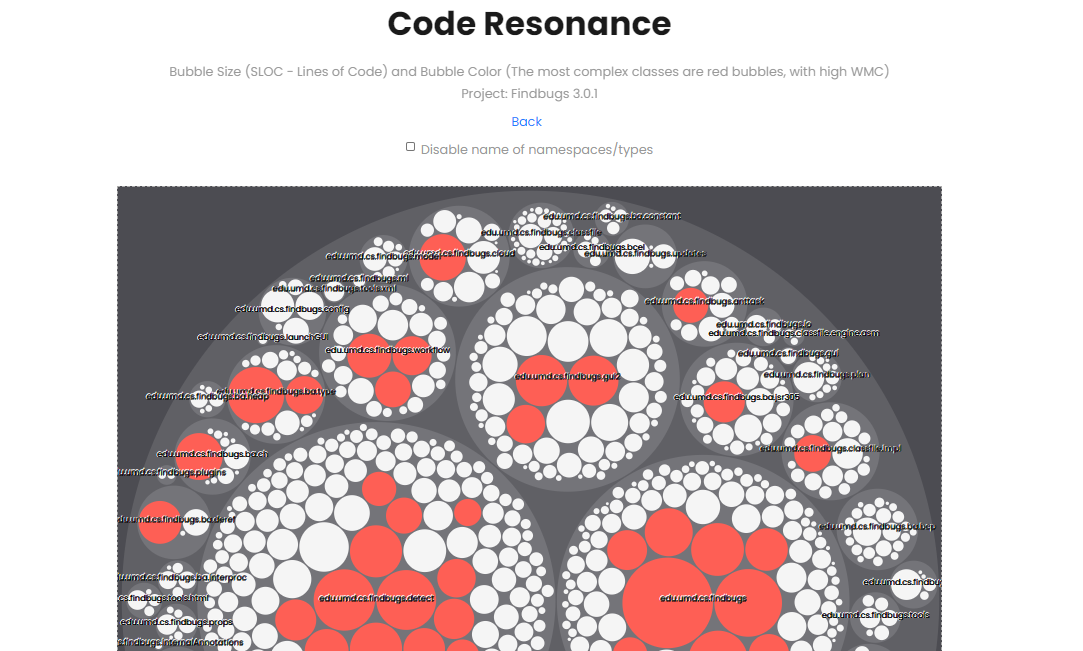}
\caption{Code resonance using \textit{DR-Tools Metric Visualization} - with this graph, we look at the distribution of the types (classes) in the namespaces (packages) and the size of the classes (bubble size). The red bubbles represent the complex classes}
\label{fig:drtools-metric-visualization-resonance}
\end{figure}

We use \textit{Javascript} and libraries like \textit{Google Chart}\footnote{More information: \url{https://developers.google.com/chart}} and \textit{D3.js}\footnote{More information: \url{https://d3js.org/}} for the graph generation.
Also, we create graphs such as bubble charts, thermometer charts, chord diagrams, bar charts, zoomable circle packing, among others, to present contextual information about the source code.
With the graphs and filters available, the team can make decisions and plan their software maintenance actions. 
For example, we present the code resonance (Figure \ref{fig:drtools-metric-visualization-resonance}). 
The views can be used by either the developer, individually, as well as by the team in technical discussion sessions or code peer/review, to plan maintenance actions.

\section{Case studies}
\label{sec:usageexamples}
%descrever metodologia
To evaluate the \textit{DR-Tools Suite}, we performed two case studies. The first case consisted of a classwork about code metrics analysis and open source projects. Using \textit{DR-Tools Suite}, students analyzed pieces of source code pointed by the tools and offered feedback. The second case consisted of behavior insights from a research analysis obtained from source code data collected in commercial software at a software house.

\subsection{Case 1 - Evaluating code metrics in Software Engineering course}

\paragraph{\textbf{Classwork Context}} 
The experiment directly involved 48 students from the Software Engineering course. Previously, students had a lesson on metrics and their importance in code analysis.
The activity consisted of defining groups with two-three students.
Each group chose open-source Java projects\footnote{The students analyzed 21 projects of different purposes and sizes, adding up to more than 3.7M LOC. The smallest project was \textit{JPacman} (2.3K LOC), and the largest project was \textit{Ghindra} (1.3M LOC)} (mostly available on \textit{GitHub}). After selecting the project, the group used \textit{DR-Tools Suite} to analyze the software in all contexts (summary, namespaces, types, methods, dependencies, and coupling). 
As students evaluated the project, they completed a survey that addressed questions about the contexts analyzed and the usability of the tool.
%% Colocar informações sobre os projetos

\paragraph{\textbf{Student's feedback}}
During the execution of the work, we got feedback on the metrics, the tool's functioning, and usability. 
Initially, students were not as familiar with metrics. In this sense, the heuristics helped them during the code analysis. With the use of the tool directing to the code, the students analyzed excerpts of the code that the tool pointed out as more problematic. Thus, the groups discussed the codes with indications of problems and also possible solutions for the code. Other points highlighted by the groups were the ease of use (directly downloading the tool, without the need for additional configuration), being a CLI tool (flexibility and combination of outputs in a single run). The ease of generating the results in an open format and being able to integrate with other tools was also a highlight. Twelve groups also produced the results and expanded the analysis with \textit{DR-Tools Metric Visualization}. The visualization allowed us to discuss and see elements (namespaces, types, methods) that were closer by combining the metrics and see if the problem was more complicated, dependent, or size related.
This feedback was valuable for us to improve the tool in many ways, especially in expanding options, including other context-relevant metrics, sorting and filtering information using the metrics. We also use this information to improve the graphics displayed in the \textit{DR-Tools Metric Visualization}.

\subsection{Case 2 - Supporting a team of developers in a code review on a commercial software}

\paragraph{\textbf{Context}}
In this case study, a team of 8 professional developers used \textit{DR-Tools Suite} in a code review session that lasted approximately 90 minutes. The project was a commercial software with over eight years in production (versions web and mobile).
Before code review session, we gave a brief presentation of the tool and the metrics used in each context.
During a session, a developer ran the tools while the others discussed the information presented. We perform a ``job shadowing" while they were analyzing the project.

\paragraph{\textbf{Insights}}
We note that developers are not as familiar with the metrics. This finding agrees with Bouwers \al \cite{Bouwers:2013}.
We saw that by combining metrics in a contextual and orderly way, with a filter mechanism, it allowed developers to analyze and make interesting inferences about the code, even before examining the source code itself. Also, they used the heuristics list to analyze the code. According to developers, the heuristics list helped to identify problematic pieces of code. 
Based on this collected information, we will propose new research defining new heuristics that use combined metrics to detect smells or code snippets that need improvement, with ranking and filtering features.
\section{Discussion}
\label{sec:discussion}

In this section, we discuss some fundamental aspects observed in case studies of \textit{DR-Tools Suite} usage. \textit{DR-Tools Suite} revealed some code analysis behaviors, and limitations, as well as impacts for researchers, instructors, and practitioners. 
In our experience, developers are familiar with CLI tools. These tools have great flexibility and allow different combinations of use. \textit{DR-Tools Metric} brings this approach, allowing the developer to make numerous combinations using the different options and results in various formats. Also, because it is a lightweight and open-source tool, the developer can integrate it with other tools in the software development workflow, like as individual code analysis process, peer-code review sessions, or even use them to integrate into a build pipeline.
By defining a set of combined metrics and some heuristics, we found that it made the process of understanding the code much easier. We note that heuristics complement the understanding of metrics.

%Impactos 
Researchers can use the \textit{DR-Tools Suite} in their research activities, empirical experiments, and studies about Software Engineering.
Instructors can use the \textit{DR-Tools Suite} to teaching classes about software quality, software maintenance/evolution, technical debt management, code smell detection, refactoring opportunities, among others.
Finally, practitioners can use the \textit{DR-Tools Suite} to plan and execute their software development activities (analysis of problematic pieces of code, refactorings, code review, among others).

%Limitações 
Currently, \textit{DR-Tools Metric} analyzes only \textit{Java} projects. However, the tool is designed to be extend to other languages and platforms. By the defined architecture, the developer only needs to implement a \textit{parser} and the \textit{corresponding visitors} for the new language, taking advantage of all the built infrastructure.
Also, we note that there is room for optimization of source code analysis, improving its performance for large projects. In the experiments but also in the development of the tools, we used some projects with hundreds of thousands of lines of code (\eg \textit{Hibernate, Spring Framework, SonarQube, Checkstyle}, and \textit{PMD}) for testing, obtaining exciting results.
Similarly, there are opportunities for improvement for \textit{DR-Tools Metric Visualization}. The creation of search engines in the charts is one of the possibilities, allowing the developer to have a more interactive tool.

\section{Conclusion}
\label{sec:conclusion}

In this paper, we introduce \textit{DR-Tools Suite}, its concepts, and tools that compose it (Section \ref{sec:drtools-suite}). The main idea is to create a set of lightweight open-source tools that help the developer in software maintenance and evolution tasks. 
We conducted two case studies (Section \ref{sec:usageexamples}). 
We intend to improve existing tools by focusing on the aspects presented in Section \ref{sec:discussion}. We observe that the use of heuristics with combined metrics and thresholds can have an interesting contribution to code comprehension.
Also, this research is a work in progress, and we intend to expand the \textit{DR-Tools Suite} with new tools, such as code smell detectors, refactoring recommenders, tools that support code review, and more.

\bibliographystyle{elsarticle-num}
\bibliography{references}

\end{document}